\documentclass[]{elsarticle}
\begin{document}
\begin{frontmatter}

\title{Solar water splitting: efficiency discussion}

\author{Jurga Juodkazyt\.{e}$^1$, Gediminas Seniutinas$^2$, Benjaminas \v{S}ebeka$^1$,\\\protect Irena Savickaja$^1$, Tadas Malinauskas$^3$, Kazimieras Badokas$^3$,\\\protect K\c{e}stutis Juodkazis$^1$, Saulius Juodkazis$^2$}
\address{$^1$~Center for Physical Sciences and
Technology, A. Go\v{s}tauto 9, LT-01108 Vilnius,
Lithuania\\\protect $^2$~Center for Micro-Photonics,  Faculty of
Science, Engineering and Technology, Swinburne University of
Technology, John St., Hawthorn, Melbourne VIC 3122,
Australia\\\protect $^3$~Institute of Applied Research, Vilnius
University, Saul\.{e}tekio Ave. 10, LT-10223 Vilnius, Lithuania}
\fntext[myfootnote]{Emails of JJ: jurgajuod@chi.lt, SJ:
sjuodkazis@swin.edu.au}

\begin{abstract}
The current state of the art in direct water splitting in
photo-electrochemical cells (PECs) is presented together with:
\emph{(i)} a case study of water splitting using a simple solar
cell with the most efficient water splitting electrodes and
\emph{(ii)} a detailed mechanism analysis. Detailed analysis of
energy balance and efficiency of solar hydrogen production are
presented. The role of hydrogen peroxide formation as an
intermediate in oxygen evolution reaction is newly revealed and
explains why an oxygen evolution is not taking place at the
thermodynamically expected 1.23~V potential.

Solar hydrogen production with electrical-to-hydrogen conversion
efficiency of 52\% is demonstrated using a simple
$\sim$0.7\%-efficient n-Si/Ni Schottky solar cell connected to a
water electrolysis cell. This case study shows that separation of
the processes of solar harvesting and electrolysis avoids
photo-electrode corrosion and utilizes optimal electrodes for
hydrogen and oxygen evolution reactions and achieves $\sim 10\%$
efficiency in light-to-hydrogen conversion with a standard 18\%
efficient household roof Si-solar cells.
\end{abstract}


\end{frontmatter}


\section{Review: materials and cell configurations}
\label{sect:intro}

A solar water splitting is decomposition of H$_{2}$O molecules
into molecular hydrogen and oxygen using solar energy. This
process is expected to become foundation of a sustainable
hydrogen-based energy economy, as it represents carbon-neutral way
to produce hydrogen gas using the most abundant renewable
resources, i.e., water and sunlight~\cite{Walter,Lewis,Parkinson}.

\subsection{Early days of solar water splitting}

From the technical point of view, there are may ways to realize
water splitting, however, the technology must be efficient and
economically viable. Comprehensive review, dedicated to the
problems of electrolysis of water on light-sensitive semiconductor
surfaces~\cite{Walter} covers the period between 1972, when the
photoelectrochemical (PEC) water splitting was first
discovered~\cite{Fujishima} and 2010. The authors survey the data
related to thermodynamics of hydrogen and oxygen evolution
reactions (HER and HOR), performance of various semiconductors,
configurations of photoelectrochemical cells, properties of
catalyst materials, influence of various structural effects on the
efficiency of the process. Values of the energy conversion
efficiency or the so-called solar-to-hydrogen efficiency,
$\eta_{STH}$, obtained for water splitting cells of different
configurations are summarized~\cite{Walter}. The values vary
between 0.01\% and 18\%, i.e., by a factor of about 2000, for PEC
cells without surface catalysts and those with catalysts and and
buried photovoltaic (PV) junctions.

In terms of a technical realisation, direct solar-to-chemical
energy conversion in PEC cell is considered to be more practical,
efficient and less expensive method of H$_2$ production compared
to electrolysis of water using PV-generated electricity, because
the former integrates light energy collection and water splitting
into one device~\cite{Turner,Sivula,Morales,Walter}. This goal,
however, poses serious material-related challenges: semiconductor
photoelectrodes should efficiently harvest solar irradiation and
drive water oxidation or reduction reactions in aqueous solutions
at a sufficient rate at current densities of (10 -
15)~mA.cm$^{-2}$ under 1 Sun illumination without degrading for a
sufficiently long period of time, i.e., more than 2000~h according
to benchmarks of US Department of Energy~\cite{Bak}. Such a set of
requirements simply cannot be met by a single material, therefore
formation of various
heterostructures~\cite{Mayer,Moniz,Choudhary,Morales,Reece,Kenney,Pijpers,Li},
tandem devices~\cite{Prevot,Ding,Gurudayal}, as well as methods
for photocorrosion protection of
semiconductors~\cite{Zhang,Kelly,Chen,Wang,Lin} have been very
intensively explored over the period of recent years in 2010 -
2015, as can be seen from the list of the referenced literature.

\subsection{Recent state-of-the-art in solar water splitting}

The highest solar-to-hydrogen efficiencies (STH), exceeding 10\%,
so far have been achieved using very expensive and unstable III-V
group semiconductors~\cite{Khaselev, Jacobsson}. Therefore the
focus of undergoing research is to find economically viable,
efficient and stable material compositions for PEC water splitting
based on earth-abundant elements. A great deal of attention is
being devoted to employment of Si photovoltaics in solar hydrogen
production, as this technology is well-developed already and, more
importantly, the price of crystalline silicon solar cells has
decreased more than 7 times in the past several years~\cite{Cox}.
Reece et al.~\cite{Reece} has reported fabrication of solar
water-splitting cells, which consist of triple junction amorphous
Si photovoltaic cell interfaced to cobalt borate and NiMoZn as
oxygen and hydrogen evolution catalysts, respectively. The highest
STH efficiency obtained was 4.7\% under 1~Sun illumination,
however, a stable operation of the device lasted only for
10~hours. Another group~\cite{Chen} has successfully stabilized
silicon photoanode by atomic layer deposition (ALD) of 2~nm thick
layer of TiO$_2$ covered with 3~nm thick iridium film for the OER
catalysis. Stable operation of the anode was observed for at least
24~hours. Similarly, ALD-grown Al-doped ZnO (20~nm) and TiO$_2$
(20~nm) protective layers modified with Pt nanoparticles have been
shown to effectively stabilize Cu$_2$O
photocathode~\cite{Paracchino2011,Paracchino2012} for H$_{2}$
production, which retained 62\% of the initial photocurrent value
after 10~hour stability test. Recently, a 40-h-long stable
operation of tandem-junction GaAs/InGaP photoanode, protected with
ALD-formed TiO$_2$ layer, in conjunction with Ni-based HER and OER
catalysts, was reported~\cite{Verlage}. Kenney et
al.~\cite{Kenney} demonstrated a record 80~h long direct water
oxidation, using n-type Si photoanodes passivated with a 2~nm
thick nickel film, which acts also as oxygen evolution catalyst.

Thus, one has to admit that significant progress has been made in
recent years in the area of engineering the semiconductor
electrodes for efficient solar water splitting, however, the
fabrication of complex heterostructures usually involves such
sophisticated techniques as
ALD~\cite{Wang,Chen,Verlage,Paracchino2011,Paracchino2012}, which
are totally incompatible with the larger area and mass production
of the photoelectrodes, if solar energy is going to be collected
on the terawatt scale. Moreover, the longevity of these
state-of-the-art photoelectrodes is still far beyond the targeted
2000~hours and the energy conversion efficiency achieves the
pursued 10\%~\cite{Bak} only when the most expensive III-V
semiconductors are used~\cite{Verlage}.

\begin{figure*}[tb]
 \centering
\includegraphics[width=11cm]{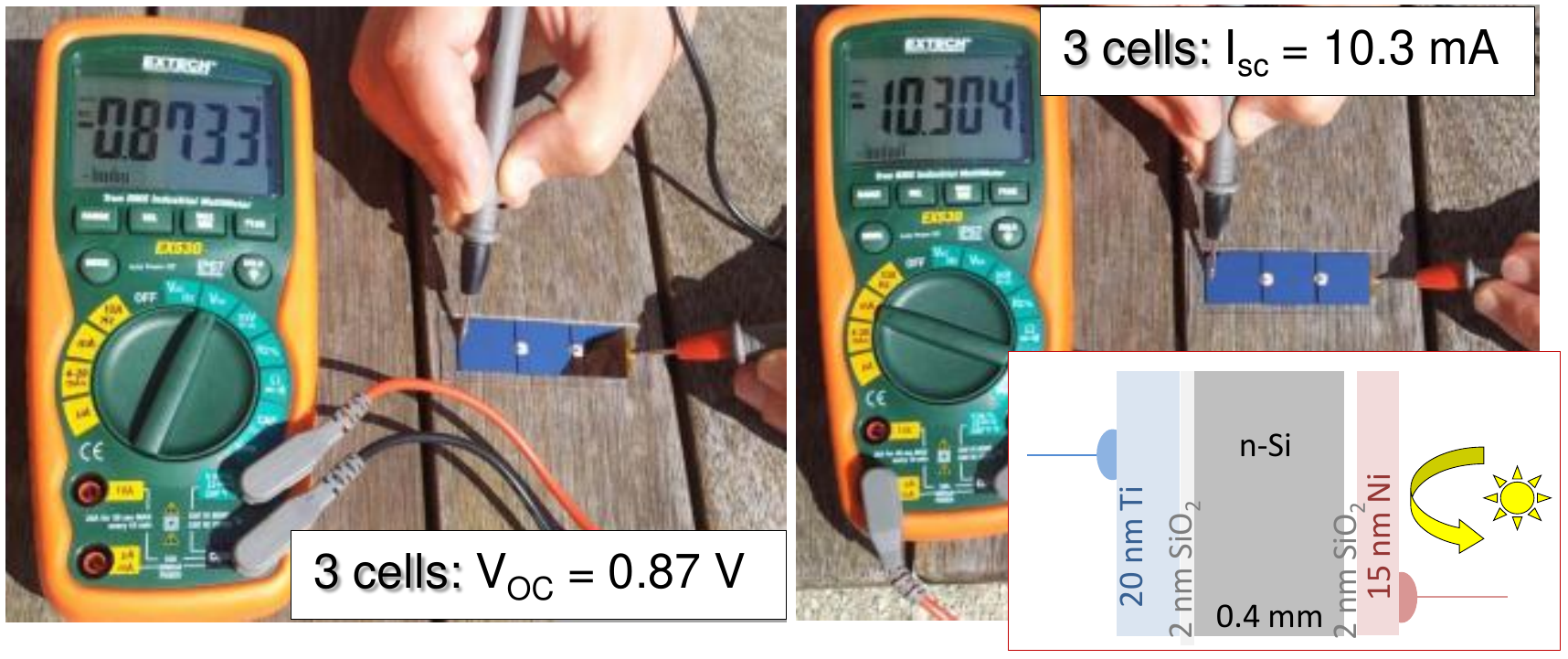}
\caption{n-Si/Ni Schottky solar cells connected in series under
natural Sun illumination (on 11.20~am, 4 March 2014, Swinburne
Uni.,  Melbourne, Australia).  The open circuit voltage, $V_{OC}$,
and short circuit current, $I_{sc}$, readings are shown on the
multimeter. Single cell area was $\sim 2\times 2$~cm$^2$. Inset
shows schematics of the layers in the n-Si/Ni Schottky diode solar
cell. The Ni and Ti were magnetron sputtered films; native SiO$_2$
of $\sim 2$~nm was on the surface of n-type Si. } \label{f-cell}
\end{figure*}

In view of the above, the so-called ``brute-force'' or
photovoltaic (PV) plus the electrolysis approach is lately being
reconsidered~\cite{10oe147,Cox}. The main advantage of this
approach is that both processes, i.e., the photovoltaic and
electrochemical energy conversion can be optimized independently.
Moreover, the problems related with photo-corrosion of
semiconductors and blocking of the light-sensitive surface with
HER or OER catalyst particles are automatically avoided. In
ref.~\cite{Peharz} $\eta_{\mathrm{STH}}= 18\%$ was obtained using
system integrating highly efficient, yet very expensive, group
III-V semiconductor solar cells, optical concentrator and polymer
electrolyte membrane electrolyzer. Quite recently~\cite{Cox} STH
efficiency exceeding 10\% has been achieved with a crystalline
silicon PV module and noble metal-free, low cost HER and OER
catalysts. The operation of the coupled system was stable for over
a week at current density of $\sim 8$~mA.cm$^{-2}$. For practical
applications an expectation of a lifetime for a solar energy
converter is dictated by solar cells and has to be 25 years. Thus,
it is evident, that in construction of efficient solar water
splitting system, a trade-off between efficiency, cost and
longevity must be achieved.

Here, we explicitly demonstrate that by separating steps of:
\emph{(1)} solar-to-electrical and \emph{(2)}
electrical-to-hydrogen conversion the highest overall
solar-to-hydrogen efficiency $\eta_{STH}$ can be achieved. As
tutorial case study, we choose a simple Si/Ni Schottky solar cell
for the process \emph{(1)} and the most efficient Pt and Ti/Ir-Ta
oxide electrodes as HER and OER catalysts for the process
\emph{(2)}. The mechanism of water splitting reactions based on
the principles of thermodynamics are revealed and the reason of
absence of water splitting at the themodynamically predicted
1.23~V potential are given.

\section{Samples and methods}

To investigate the efficiency of water splitting a simple setup
was made using a Schottky solar cell and connecting it to a water
splitting cell.

\subsection{Fabrication of Schottky solar cells}

Solar cell was made on a $2\times 2$~cm$^{2}$ n-type Si wafer
(0.4~mm thick) by sputtering 15-nm-thick Ni layer for the front
(illuminated) side and a 20~nm back side Ti contact
(Fig.~\ref{f-cell}), the photoanode design used for the water
oxidation~\cite{Kenney}. Metal coatings were sputtered using AXXIS
(JKLesker) physical vapor deposition setup. Si (100) wafers were
n-type of a 0.3-0.5~$\Omega$.cm resistivity (University Wafers).
The values of the open-circuit voltage, $V_{oc}$, and
short-circuit current, $I_{sc}$, of three cells connected in
series and measured with multimeter under natural outdoor
illumination conditions were 0.87~V and 10.3~mA, respectively
(Fig.~\ref{f-cell}).

In order to efficiently drive water-splitting reactions, the
voltage applied on water electrolysis cell should exceed
1.8~V~\cite{Prevot}, therefore we used the battery of nine
Schottky solar cells in our experiments. A home designed solar
simulator was used for experiments of solar-to-hydrogen
conversion. A high intensity discharge Xe-lamp with 6000~K
spectrum was used to illuminate solar cells using a collimating
lens and was calibrated with a silicon diode to have approximately
50~mW.cm$^{-2}$ irradiance (0.5 Sun) at 30~cm from the lamp over
the area of $5\times 5$~cm$^2$.

Current-voltage $I-V$ characteristics of the battery of Schottky
solar cells were measured using Keithley 2400 SourceMeter. The
$I-V$ response upon illumination was measured using the same
illumination device and conditions as in the electrochemical
experiments, namely, the intensity of illumination was
50~mW.cm$^{-2}$ (0.5 Sun).

\subsection{Electrochemical measurements}

The PV-assisted water electrolysis experiments were performed in
glass cell of 30~cm$^{3}$ volume using Pt cathode ($S_c =
1~$cm$^{2}$) for HER and dimensionally stable (DSA) titanium anode
($S_a = 1$~cm$^{2}$) with catalytic layer of mixed iridium and
tantalum oxides (Elade Technology Co. Ltd, China) for OER.
Electrolyte was aqueous solution of 0.5~M Na$_{2}$SO$_4$ and 0.1~M
KOH. Reagents of analytical grade and deionized water were used to
prepare the solution. The operating voltage, $V_{op}$, and the
operating current density, $i_{op}$, in the water electrolysis
cell between the anode and cathode, connected through 1~$\Omega$
resistance, were measured using potentiostat-galvanostat PI-50-1
connected to computer through an analogue-digital converter
interface.

Cyclic voltammetry was used to test the individual performance of
the above described anode and cathode in the same electrolyte
solution. The experiments were performed using
potentiostat/galvanostat AUTOLAB 302 and three-electrode
electrochemical cell. Pt plate served as counter electrode (CE)
and reversible hydrogen electrode in working solution (RHE) was
used as the reference. Potential values in the text refer to the
RHE scale, unless noted otherwise. The conversion between RHE and
standard hydrogen electrode (SHE) scales can be done according to
equation: $E_{\mathrm{vs.SHE}} = E_{\mathrm{vs.RHE}} -
0.059\mathrm{pH}$. All experiments were performed at room
conditions.

\section{Results: water splitting}\label{sect:schot}

Efficiency analysis of the two intermediate steps in
solar-to-hydrogen conversion:  \emph{(1)} solar-to-electrical and
\emph{(2)} electrical-to-hydrogen are tested using simple solar
cell and water splitting cell.

\subsection{I-V characteristics of Schottky solar cells}
\begin{figure*} \centering
\includegraphics[width=13cm]{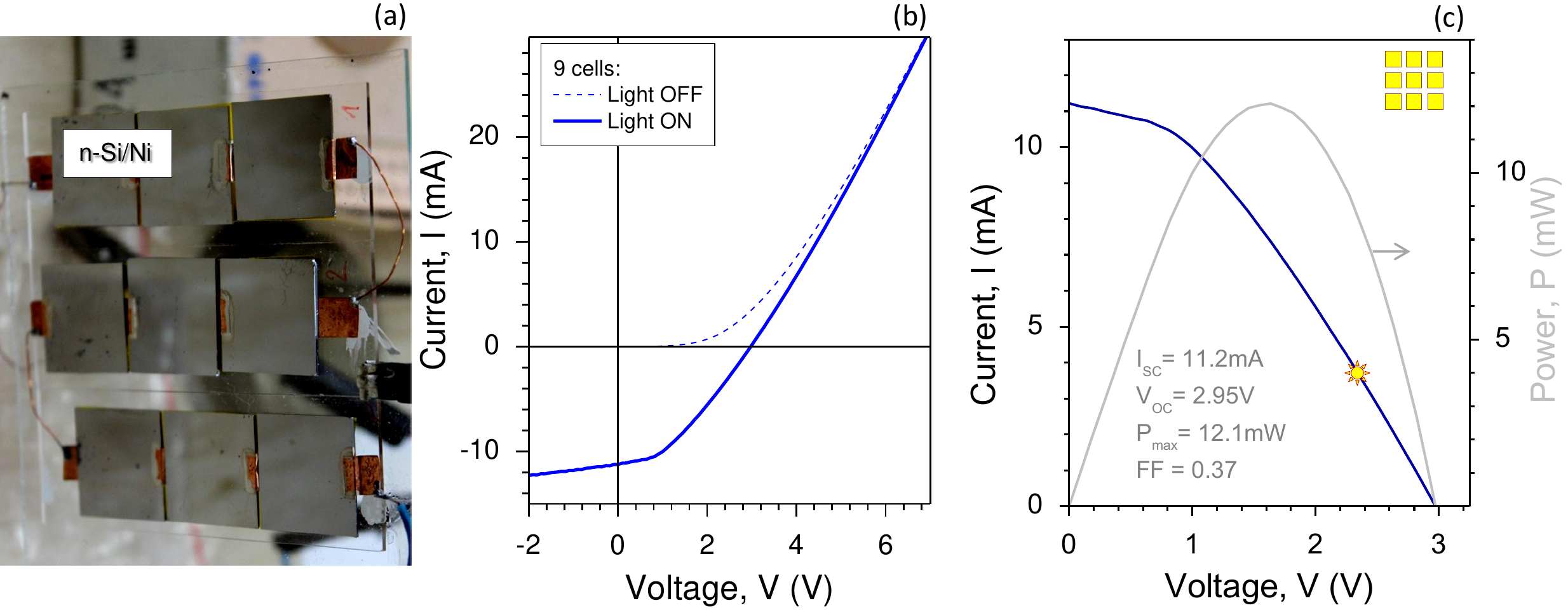}
\caption{(a) Solar cell battery assembled from cells shown in
Fig.~\ref{f-cell} used in experiments of water splitting; (b) I-V
plot for nine cells connected in series in dark and under
artificial 0.5 Sun illumination. (c) Determination of the fill
factor, $FF$, of the battery of nine cells under an artificial 0.5
Sun illumination: $V_{OC} = 2.95$~V, $I_{sc} = 11.2$~mA, $P_{max}
= 12.1$~mW yielding $FF = 0.37$. The $\otimes$ marker shows the
working point voltage in the case of solar water splitting using
solar cell battery coupled with water electrolysis
cell.}\label{f2jurgos}
\end{figure*}

Figure~\ref{f2jurgos}(a) shows the photo of battery of n-Si/Ni
Schottky solar cells connected in series and mounted on a slice
glass, which was used in water-splitting experiments.
Figure~\ref{f2jurgos} (b) shows the current-voltage, I-V,
characteristics of this battery in dark and under artificial 0.5
Sun illumination with a typical voltammetric response of a
Schottky photodiode. Figure~\ref{f2jurgos}(c) illustrates the
determination of the fill factor for the battery of 9 cells, $FF =
P_{max}/(V_{oc}\times I_{sc})$, where $V_{oc}$ is the open-circuit
voltage, $I_{sc})$ is short-circuit current, and $P_{max}$ is the
maximum power, measured under 0.5 Sun artificial illumination. The
$FF$ value was further used for the estimation of the
light-to-electric (photovoltaic) energy conversion efficiency,
$\eta_{~\mathrm{PV}}$, which is defined as the ratio between
useful power output $P_{out}$, and total light power input $P_{in}
= S\times I_{l}$ factored over the solar harvesting area
$S$:
\begin{equation}\label{e1a}
\eta_{PV} \equiv \frac{P_{max}}{P_{in}} = \frac{ I_{sc}
V_{OC}\times FF}{S\times I_{l}},
\end{equation}
where light intensity was $I_l = 50~$mW/cm$^2$ or 0.5 Sun used in
experiments and as harvested over $S = 9\times (2\times 2) =
36$~cm$^{2}$ area.

Rather small values of $FF$ = 0.37 and $\eta_{PV} \simeq 0.68$~\%
were obtained. As one can see from Fig.~\ref{f2jurgos}(c), the
current at maximum power point is $\sim 7$~mA. Having in mind
large area of illuminated surface as well as low value of fill
factor, one has to admit that n-Si/Ni Schottky solar cell is not
an optimal choice, however, its operation in air atmosphere is
stable; as a photoanode in direct water splitting the same
Schottky barrier has lifetime $< 100$~h~\cite{Kenney}. This simple
solar cell is used here to reveal factors defining the
efficiencies of an overall solar-to-hydrogen conversion and to
show pathway to the highest values. From the slope of I-V curve in
dark (Fig.~\ref{f2jurgos}(b)) the resistance of the battery of
9-cells in series was evaluated as $\approx 130~\Omega$ or
$R_{1cell} = 14~\Omega$ due to a native oxide layer.

\subsection{Water splitting: choice of electrodes}

Pt cathode was chosen for the hydrogen evolution reaction. Though
Pt is expensive material, it is the best and stable HER catalyst
in both acidic and alkaline solutions~\cite{McCrory}. Commercially
available dimensionally stable titanium-supported iridium/tantalum
oxide-based anode was tested for oxygen evolution reaction. To
evaluate the overpotentials of hydrogen and oxygen evolution
reactions, cyclic voltammograms of the chosen electrodes were
recorded in the solution of 0.5~M Na$_{2}$SO$_4$ and 0.1~M KOH and
are shown in Fig.~\ref{f4jurgos}. The areas negative and positive
to vertical lines at $E = 0$~V and $E = 1.23$~V, correspond to
overpotentials of HER and OER, respectively. It should be noted,
however, that ohmic potential drop due to resistance of the
electrolyte is not taken into consideration in
Fig.~\ref{f4jurgos}, however it has a minor effect.

\subsection{Water splitting powered by Schottky solar cell}

\begin{figure}[tb]\centering
\includegraphics[width=9.0cm]{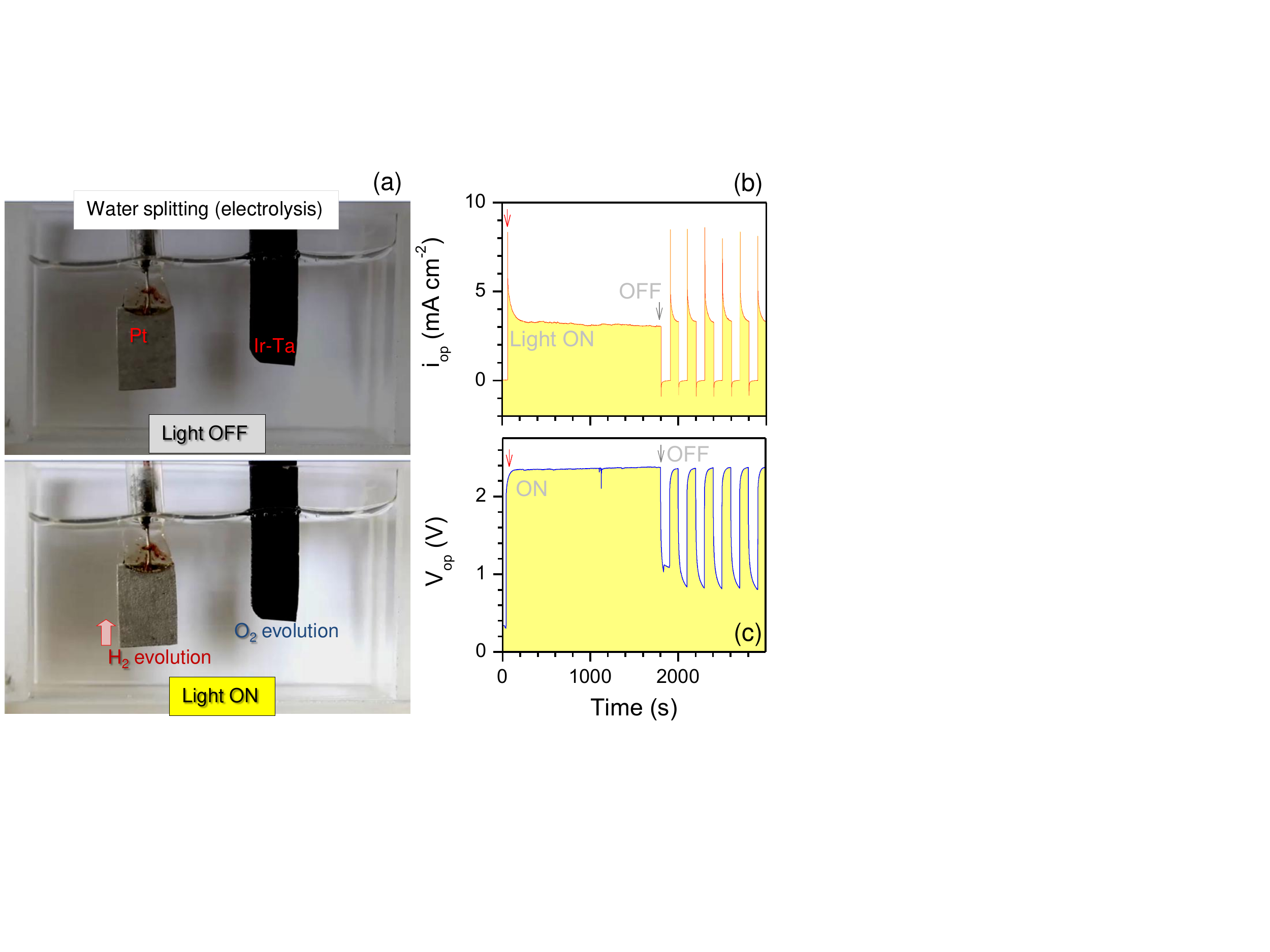}
\caption{(a) Photo of water photoelectrolysis cell with Pt cathode
for H$_2$ evolution and Ti/Ir-Ta anode for O$_2$ evolution. See a
supplement movie of solar water splitting. (b) Variation of
photocurrent in water photoelectrolysis cell upon switching ON/OFF
the illumination on n-Si/Ni Schottky solar cell battery
(Fig.~\ref{f2jurgos} (a)). The anode and cathode area in
experiments of water splitting was $S_{a,c}\sim
1$~cm$^2$.}\label{f-bubbles}
\end{figure}

We coupled photovoltaic and electrochemical systems in the most
simple way, i.e., by wiring the positive and negative terminals of
solar cell battery (Fig.~\ref{f2jurgos}(a)) to anode and cathode
of the water electrolysis cell, respectively. Evolution of gases
on both electrodes was recognized almost instantaneously by
formation of micro-bubbles (see, the supplement movie) after
switching the light illumination (Fig.~\ref{f-bubbles}(a)).

\begin{figure}\centering
\includegraphics[width=6.0cm]{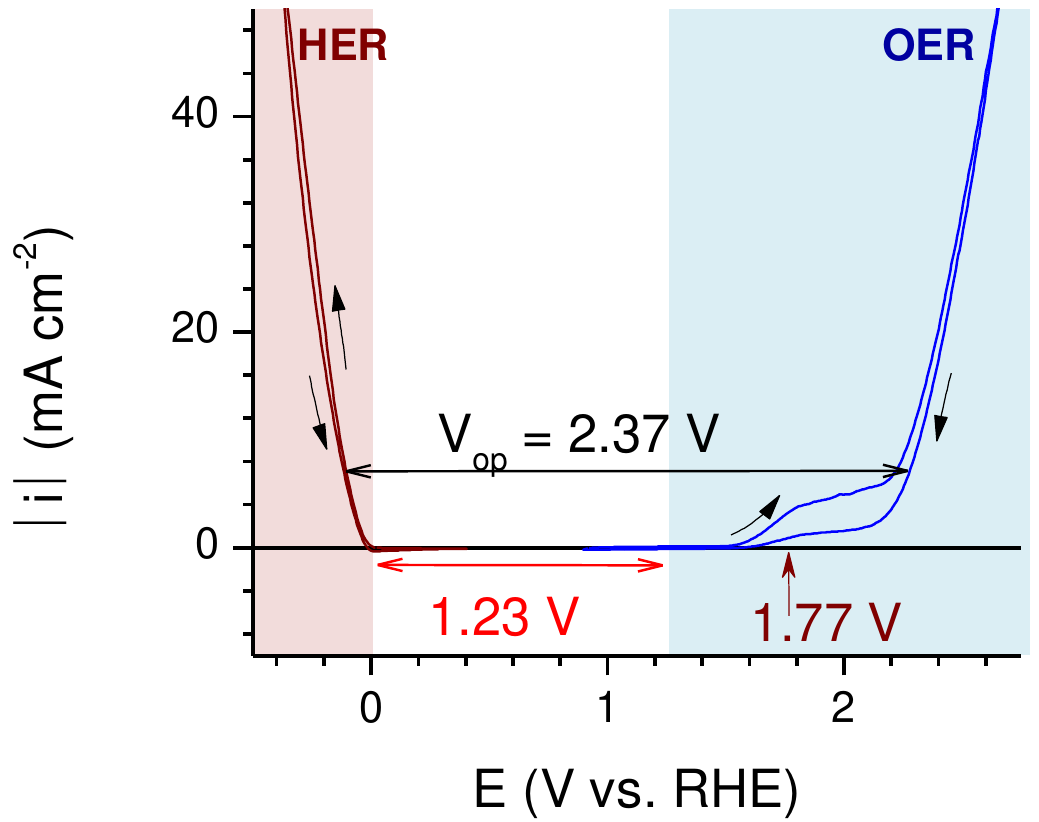}
\caption{Cyclic voltammograms of Pt cathode and Ti/(Ir-Ta) anode
in HER and OER regions, respectively, solution 0.5~M
Na$_{2}$SO$_4$ + 0.1~M KOH,  potential scan rate 10~mV.s$^{-1}$.
The arrow at $E^0 = 1.77$~V marks the standard potential of
H$_{2}$O$_{2}$ formation which is suggested to be the threshold
potential in oxygen evolution reaction (see,
Eqns.~\ref{e2}-\ref{e3}). Electrode anode and cathode area in
experiments of water splitting was $S_{a,c}\sim
1$~cm$^2$.}\label{f4jurgos}
\end{figure}

Initially, upon switching on the light, the operating current
density, $i_{op}$, in the water electrolysis cell jumped to $\sim
8$~mA.cm$^{-2}$. Then a stationary value of $\sim 3$~mA.cm$^{-2}$
was attained within half an hour (Fig.~\ref{f-bubbles}(b).
Chopping of illumination resulted in reproducible current density
transients. The variation of the operating voltage $V_{op}$ of the
cell during electrolysis under constant and chopped illumination
is shown in Fig.~\ref{f-bubbles}(c). The stationary value was
$V_{op} = 2.37$~V. Given such operating voltage, the water
splitting reactions, in accordance with Fig.~\ref{f4jurgos}, could
be driven at the rate of $\sim 7$~mA.cm$^{-2}$, however, the
actual values are dictated by the I-V characteristic (low $FF$) of
the PV power source (see the working point of coupled system
indicated in Fig.~\ref{f2jurgos} (c)) and the resistance of the
whole circuit.

The overall or solar-to-fuel efficiency of the system can be
calculated according to the following equation:
\begin{equation}\label{e-ef}
\eta_{STH} \equiv \eta_{PV}\times \eta_{ec}\times \eta_{co},
\end{equation}
\noindent where $\eta_{PV}$ is the efficiency of light-to-electric
energy conversion, $\eta_{ec}$ is the efficiency of
electric-to-chemical energy conversion and $\eta_{co}$ is the
efficiency of coupling of the two systems~\cite{Winkler}. The
efficiency of the electrochemical water splitting is:
\begin{equation}\label{e1}
\eta_{ec} = [1.23~\mathrm{V}\times I_{op}]/[I_{op}V_{op}] \simeq
0.52~~~(\mathrm{or}~52\%),
\end{equation}
\noindent where 1.23~V corresponds to the thermodynamic potential
of water splitting reaction, $I_{op} = 3$~mA is the operating
current through the water electrolysis cell at $V_{op} = 2.37$~V.
The coupling efficiency can be evaluated on the basis of the
following reasoning. The photovoltaic efficiency of the solar cell
battery at the operating point is $\eta_{op} =
I_{op}V_{op}/[S\times I_{l}] \simeq 0.4\%$, while the maximum
$\eta_{PV}$ is 0.68\% (see Eqn.~\ref{e1a}). Consequently, the
coupling efficiency can be calculated according to: $\eta_{co} =
\eta_{op}/\eta_{PV} \simeq 0.59$ (59\%). Hence, the overall
cumulative efficiency of solar-to-hydrogen conversion is
$\eta_{STH} \simeq 0.68\times 10^{-2}\times 0.59\times 0.52 =
0.21\%$.

In order to achieve higher values of energy conversion efficiency,
more efficient photovoltaics should be used. Efficiency of the
electrolysis is not a limiting factor in this case. The coupling
efficiency of 59\% in this case study is due to an electrical
impedance mismatching and could be further improved. Despite low
value of $\eta_{STH}$, such method of solar hydrogen production by
separating \emph{(1)} solar-to-electrical and \emph{(2)}
electrical-to-hydrogen steps could be competitive considering its
long term stability.

\section{Discussion}
\label{sect:disco}

The presented tutorial demonstration of solar hydrogen production
with $\eta_{STH} \simeq 0.21\%$ efficiency using a low efficiency
$\eta_{PV} \simeq 0.68\%$ solar cell is analysed next for more
practical implementation and analysis of fundamental limitations.

\subsection{Some mechanistic aspects of water splitting reactions}

From the thermodynamic point of view, the voltage of only 1.23~V
should be should be sufficient to \emph{electrolytically} split
water into O$_{2}$ and H$_{2}$, since
$E^{0}_{\mathrm{O}_{2}/\mathrm{H}_{2}\mathrm{O}} = 1.23$~V and
$E^{0}_{\mathrm{H}^{+}/\mathrm{H}_{2}} = 0$~V under standard
conditions. However, as it can be explicitly seen in
Fig.~\ref{f4jurgos}, at $\Delta E = 1.23$~V the current density $i
\approx 0$. It is well known that modern alkaline electrolyzers
operate at voltages exceeding 1.8~V~\cite{Prevot,Walter} and
appreciable rate of HER and OER is achieved at 2.0 - 2.5~V. It is
also well-known that HER can be catalyzed at the overpotentials
much lower than those of OER~\cite{Winkler,McCrory}. The mechanism
of hydrogen evolution reaction (HER) on Pt electrode mediated by
molecular hydrogen ion has been described
earlier~\cite{14ass13,11ass743,13spie88220S}. The mechanism of
oxygen evolution reaction (OER), taking place on the oxidized
metal electrode surface, should involve formation of hydrogen
peroxide as intermediate, followed by its oxidation to O$_{2}$
according to equations as follows:
\begin{equation}\label{e2}
2\mathrm{H}_{2}\mathrm{O} - 2e^{-} \Leftrightarrow
\mathrm{H}_{2}\mathrm{O}_{2} + 2\mathrm{H}^{+},~~~~~~~~ E^{0} =
1.77~V,
\end{equation}
\begin{equation}\label{e3}
\mathrm{H}_{2}\mathrm{O}_{2} - 2e^{-} \Leftrightarrow
\mathrm{O}_{2} + 2\mathrm{H}^{+},~~~~~~~~ E^{0} = 0.68~V.
\end{equation}
\noindent Such layout of the $E^{0}$ values of reactions~\ref{e2}
and \ref{e3} means that, once formed, hydrogen peroxide is
spontaneously oxidized to O$_{2}$. Sum of these two reactions
divided by 2 gives the formal equation of the OER:
\begin{equation}\label{e4}
\mathrm{H}_{2}\mathrm{O} - 2e^{-} \Leftrightarrow
1/2\mathrm{O}_{2} + 2\mathrm{H}^{+},~~~~~~ E^{0} = 1.23~V,
\end{equation}
\noindent while the process, actually, does not follow this path
(see OER region in Fig.~\ref{f4jurgos}). Therefore, reaction
(\ref{e2}) with $E^{0} = 1.77$~V should be considered to be the
main energetic barrier in H$_{2}$O splitting process. Since
thermodynamically oxidation of water is allowed at 1.23~V, it can
be depolarized in the range between 1.23~V and 1.77~V with a help
of suitable electrocatalysts. The mechanism of electrocatalytic
OER taking place in the case of Ru and Ni electrodes and involving
the formation of metal surface peroxo species has been
proposed~\cite{MesRu,MesNi}.

In the case of \emph{photolytic} water splitting on the
semiconductor surface, the photon with an energy of 1.23~eV
($\lambda \approx 1~\mu$m) is able to break one O-H bond in
H$_{2}$O molecule. So, to break both bonds, two such photons are
needed, i.e., thermodynamic work equivalent to $\sim 2.46$~eV
should be performed. The same amount of energy is gained when one
mole of H$_{2}$ gas is burned in O$_{2}$ atmosphere. The same work
would be performed by two photons with the energies of $\geq 1.77
\mathrm{eV} + 0.68$~eV, or one photon with the energy of $\geq
2.46$~eV. Thus, in terms of energy, the expenses in both cases,
i.e. electrolytic and photolytic water splitting, are the same.

\subsection{Approaching theoretical limits}

An understanding of fundamental mechanisms of solar-to-electrical
energy conversion has recently resulted in fast progress towards
theoretical Shockely-Queisser 33.5\% limit of solar energy
harvesting. It has been demonstrated~\cite{YablonovitchSPIE}, that
solar-to-electrical conversion up to 28.8\% can be achieved using
direct bandgap GaAs material with inherently large luminescence
yield of 99.7\%, due to augmented $V_{OC} = 1.12$~V. This
corresponded to a 2.4\% increase made in one year from the
previous record in GaAs; a 30\% efficiency seems is in reach in
the near future. The record high efficiency will be achieved in
few micrometers thin solar cell. In the solar-to-hydrogen
conversion, the most efficient method is to separate
solar-to-electrical and electrical-to-chemical (via electrolysis)
parts as shown in this study. Understanding of HER and OER
mechanisms at molecular level is essential in search for the most
efficient water splitting catalysts. Storing of solar energy in
the form of chemical bonds (hydrogen) is providing a method to
solve inherent inefficiencies in the day-night and summer-winter
solar energy cycling.

The current hosehold PV panels are working at $\eta_{PV} \simeq
18\%$ and the water electrolysis efficiencies of $\eta_{ec} \simeq
60\%$ are typical~\cite{Bard}. For the current state-of-the-art,
the overall efficiency of solar hydrogen production and ideal
$\eta_{co} = 100\%$ (achievable via micro-invertor control) is
expected to approach (Eqn.~\ref{e-ef}): $\eta_{STH} = 0.18\times
0.6 \simeq 0.11$ (11\%). This estimate shows that separation of
solar-to-electric and electric-to-chemical processes and
increasing their efficiencies $\eta_{PV}$ and $\eta_{ec}$ is an
unrivaled approach for future solar hydrogen technology. The
efficiencies achievable today are more that one order of magnitude
larger as compared with direct light-to-hydrogen conversion
efficiencies. Moreover, those direct water splitting solutions
have not reached required long term stabilities of the process
(years of operation are required).

For practical use of solar hydrogen in future, sea water has to be
used. This would further complicate electrode stability for the
direct light-to-hydrogen conversion schemes which are even not
currently researched. Desalination of water is an expensive
process as it uses large amounts of energy, therefore new anode
materials should be designed, that would preferentially evolve
oxygen instead of chlorine from seawater~\cite{Parkinson}.
Manganese-tungsten oxides deposited on iridium-oxide coated
titanium have been shown to suppress chlorine evolution and
enhance the OER in seawater electrolysis~\cite{Izumiya}.
Ti-supported catalytic layer of mixed Ru and Ni oxides has been
suggested as the electrode material suitable for obtaining pure
oxygen from sea water alkalized to pH $\geq 14$~\cite{MesRu}.

\section{Conclusions}
\label{sect:conc}

Critical analysis for available literature on solar hydrogen
generation is presented with a case study analysis using a simple
solar cell and performing water splitting in an electrolysis cell.

It is demonstrated that a very simple battery of n-Si/Ni Schottky
type solar cells can be used for the solar hydrogen generation
with $\sim$0.2\% efficiency, when the processes of solar energy
harvesting and electrolysis are separated. Low efficiency of
light-to-electric energy conversion, $\eta_{PV}$ = 0.68\%, is
shown to be a limiting factor in the overall process of
light-to-chemical energy conversion in this analysed case.

It is suggested that oxidation of water to hydrogen peroxide with
$E^{0}_{H_{2}O_{2}/2H_{2}O} = 1.77$~V should be considered as the
main energetic barrier in oxygen evolution reaction. Even with
such low efficiency $\eta_{STH} = 0.2$~\%, the described way of
solar hydrogen production could be competitive considering the
long term stability of operation of the coupled systems.
Directions for efficient and practical solutions in solar hydrogen
production based on mechanism and considering current and future
PV technology are presented.

\small\section*{Acknowledgments} SJ and GS are grateful for the
partial support via the Australian Research Council DP130101205
Discovery project and to Swinburne University for the start up
funding of the Nanotechnology facility by a strategic
infrastructure grant.

\section*{References}

\small
\begin{thebibliography}{10}
\expandafter\ifx\csname url\endcsname\relax
  \def\url#1{\texttt{#1}}\fi
\expandafter\ifx\csname
urlprefix\endcsname\relax\def\urlprefix{URL }\fi
\expandafter\ifx\csname href\endcsname\relax
  \def\href#1#2{#2} \def\path#1{#1}\fi

\bibitem{Walter}
M.~G. Walter, E.~L. Warren, J.~R. McKone, S.~W. Boettcher, Q.~Mi,
E.~A.
  Santori, N.~S. Lewis, Solar water splitting cells, Chem. Rev. 110 (2010)
  6446--6473.

\bibitem{Lewis}
N.~S. Lewis, D.~G. Nocera, Powering the planet: Chemical
challenges in solar
  energy utilization, PNAS 103~(43) (2006) 15729 -- 15735.

\bibitem{Parkinson}
B.~A. Parkinson, J.~Turner, Photoelectrochemical Water Splitting:
Materials,
  Processes and Architectures, Vol.~9 of Ener. Environm. Series, RSC
  Publishing, 2013.

\bibitem{Fujishima}
A.~Fujishima, K.~Honda, Electrochemical photolysis of water at a
semiconductor
  electrode, Nature 238 (1972) 37--38.

\bibitem{Turner}
J.~A. Turner, A nickel finish protects silicon photoanodes for
water splitting,
  Science 342 (2013) 811--812.

\bibitem{Sivula}
K.~Sivula, F.~L. Formal, M.~Gratzel, Solar water splitting:
Progress using
  hematite {alpha-Fe$_2$O$_3$}, ChemSusChem 4 (2011) 432--449.

\bibitem{Morales}
C.~G. Morales-Guio, S.~D. Tilley, H.~Vrubel, M.~Gratzel, X.~Hu,
Hydrogen
  evolution from a copper(i) oxide photocathode coated with an amorphous
  molybdenum sulphide catalyst, Nature Comm. 5 (2014) 3059.

\bibitem{Bak}
T.~Bak, J.~Nowotny, M.~Rekas, C.~C. Sorrell, Photo-electrochemical
hydrogen
  generation from water using solar energy. materials-related aspects., Int. J.
  Hydrogen Ener. 27 (2002) 991--1022.

\bibitem{Mayer}
M.~T. Mayer, Y.~Lin, G.~Yuan, D.~Wang, Forming heterojunctions at
the nanoscale
  for improved photoelectrochemical water splitting by semiconductor materials:
  Case studies on hematite, Acc. Chem. Res. 46~(7) (2013) 1558--1566.

\bibitem{Moniz}
S.~J.~A. Moniz, S.~A. Shevlin, D.~J. Martin, Z.-X. Guo, J.~Tang,
Visible -
  light driven heterojunction photocatalysts for water splitting - a critical
  review, Energy Environ. Sci. 8 (2015) 731 -- 759.

\bibitem{Choudhary}
S.~Choudhary, S.~Upadhyay, P.~Kumar, N.~Singh, V.~R. Satsangi,
R.~Shrivastav,
  S.~Dass, Nanostructured bilayered thin films in photoelectrochemical water
  splitting - a review, Int. J. Hydrogen Ener. 37 (2012) 18713--18730.

\bibitem{Reece}
S.~Y. Reece, J.~A. Hamel, K.~Sung, T.~D. Jarvi, A.~J. Esswein,
J.~J.~H.
  Pijpers, D.~G. Nocera, Wireless solar water splitting using silicon-based
  semiconductors and earth-abundant catalysts, Science 334 (2011) 645--648.

\bibitem{Kenney}
M.~J. Kenney, M.~Gong, Y.~Li, J.~Z. Wu, J.~Feng, M.~Lanza, H.~Dai,
  High-performance silicon photoanodes passivated with ultrathin nickel films
  for water oxidation, Science 342 (2013) 836--840.

\bibitem{Pijpers}
J.~J.~H. Pijpers, M.~T. Winkler, Y.~Surendranath, T.~Buonassisi,
D.~G. Nocera,
  Light-induced water oxidation at silicon electrodes functionalized with a
  cobalt oxygen-evolving catalyst, PNAS 108 (2011) 10056--10061.

\bibitem{Li}
Z.~Li, W.~Luo, M.~Zhang, J.~Feng, Z.~Zou, Photoelectrochemical
cells for solar
  hydrogen production: current state of promising photoelectrodes, methods to
  improve their properties, and outlook, Energy Environ. Sci. 6 (2013) 347 --
  370.

\bibitem{Prevot}
M.~S. Prevot, K.~Sivula, Photoelectrochemical tandem cells for
solar water
  splitting, J. Phys. Chem. C 117 (2013) 17879--17893.

\bibitem{Ding}
C.~Ding, W.~Qin, N.~Wang, G.~Liu, Z.~Wang, P.~Yan, J.~Shi, C.~Li,
  Solar-to-hydrogen efficiency exceeding 2.5 percent achieved for overall water
  splitting with an all earth-abundant dual photoelectrode, Phys. Chem. Chem.
  Phys. 16 (2014) 15608--15614.

\bibitem{Gurudayal}
Gurudayal, D.~Sabba, M.~H. Kumar, L.~H. Wong, J.~Barber,
M.~Gratzel,
  N.~Mathews, Perovskite-hematite tandem cells for efficient overall solar
  driven water splitting, Nano Letters 15 (2015) 3833--3839.

\bibitem{Zhang}
Z.~Zhang, R.~Dua, L.~Zhang, H.~Zhu, H.~Zhang, P.~Wang,
Carbon-layer-protected
  cuprous oxide nanowire arrays for efficient water reduction, ACS Nano 7
  (2013) 1709--1717.

\bibitem{Kelly}
N.~A. Kelly, T.~L. Gibson, Design and characterization of a robust
  photoelectrochemical device to generate hydrogen using solar water splitting,
  Inter. J. Hydrogen Ener. 31 (2006) 1658--1673.

\bibitem{Chen}
Y.~W. Chen, J.~D. Prange, S.~Duhnen, Y.~Park, M.~Gunji, C.~E.~D.
Chidsey, P.~C.
  McIntyre, Atomic layer-deposited tunnel oxide stabilizes silicon photoanodes
  for water oxidation, Nature Mater. 10 (2011) 539--544.

\bibitem{Wang}
T.~Wang, Z.~Luo, C.~Li, J.~Gong, Controllable fabrication of
nanostructured
  materials for photoelectrochemical water splitting via atomic layer
  deposition, Chem. Soc. Rev. 43 (2014) 7469--7484.

\bibitem{Lin}
Y.~Lin, Y.~Xu, M.~T. Mayer, Z.~I. Simpson, G.~McMahon, S.~Zhou,
D.~Wang, Growth
  of p-type hematite by atomic layer deposition and its utilization for
  improved solar water spliting, J. Am. Chem. Soc. 134 (2012) 5508--5511.

\bibitem{Khaselev}
O.~Khaselev, A.~Bansal, J.~A. Turner, High-efficiency integrated
multijunction
  photovoltaic/electrolysis systems for hydrogen production, Int. J. Hydrogen
  Ener. 26 (2001) 127--132.

\bibitem{Jacobsson}
T.~J. Jacobsson, V.~Fjallstrom, M.~Sahlberg, M.~Edoff,
T.~Edvinsson, A
  monolithic device for solar water splitting based on series interconnected
  thin film absorbers reaching over 10 {\%} solar-to-hydrogen efficiency,
  Energy Environ. Sci. 6 (2013) 3767--3683.

\bibitem{Cox}
C.~R. Cox, J.~Z. Lee, D.~G. Nocera, T.~Buonassisi, Ten-percent
solar-to-fuel
  conversion with nonprecious materials, PNAS 111~(39) (2014) 14057--14061.

\bibitem{Paracchino2011}
A.~Paracchino, V.~Laporte, K.~Sivula, M.~Gratzel, E.~Thimsen,
Highly active
  oxide photocathode for photoelectrochemical water reduction, Nature Mater. 10
  (2011) 456--461.

\bibitem{Paracchino2012}
A.~Paracchino, N.~Mathews, T.~Hisatomi, M.~Stefik, S.~D. Tilley,
M.~Gratzel,
  Ultrathin films on copper(i) oxide water splitting photcathodes: a study on
  performance and stability, Ener. Environ. Sci. 5 (2012) 8673--8681.

\bibitem{Verlage}
E.~Verlage, S.~Hu, R.~Liu, R.~J.~R. Jones, K.~Sun, C.~Xiang, N.~S.
Lewis, H.~A.
  Atwater, A monolithically integrated, intrinsically safe, 10 percent
  efficient, solar-driven water-splitting system based on active, stable
  earth-abundant electrocatalysts in conjunction with tandem {III–V} light
  absorbers protected by amorphous {TiO}$_2$ films, Ener. Envir. Sci. Advance
  Article (2015) DOI: 10.1039/C5EE01786F.

\bibitem{10oe147}
K.~Juodkazis, J.~Juodkazyt\.{e}, P.~Kalinauskas, E.~Jelmakas,
S.~Juodkazis,
  Photoelectrolysis of water: Solar hydrogen - achievements and perspectives,
  Opt. Express: energy express 18 (2010) A147--A160.

\bibitem{Peharz}
G.~Peharz, F.~Dimroth, U.~Witstadt, Solar hydrogen production by
water
  splitting with a conversion efficiency of 18 percent, Int. J. Hydrogen Ener.
  32 (2007) 3248--3252.

\bibitem{McCrory}
C.~McCrory, S.~Jung, I.~M. Ferrer, S.~M. Chatman, J.~C. Peters,
T.~F.
  Jaramillo, Benchmarking hydrogen evolving reaction and oxygen evolving
  reaction electrocatalysts for solar water splitting devices, J. Am. Chem.
  Soc. 137 (2015) 4347--4357.

\bibitem{Winkler}
M.~T. Winkler, C.~R. Cox, D.~G. Nocera, T.~Buonassisi, Modeling
integrated
  photovoltaic - electrochemical devices using steady - state equivalent
  circuits, PNAS (2013) E1076 -- E1082.

\bibitem{14ass13}
K.~Juodkazis, J.~Juodkazyt\.{e}, B.~\v{S}ebeka, S.~Juodkazis,
Reversible
  hydrogen evolution and oxidation on {Pt} electrode mediated by molecular ion,
  Appl. Surf. Sci. 290 (2014) 13--17.

\bibitem{11ass743}
K.~Juodkazis, J.~Juodkazyt\.{e}, A.~Grigucevi\v{c}ien\.{e},
S.~Juodkazis,
  Hydrogen species within the metals: role of molecular hydrogen ion {H$_2^+$},
  Appl. Surf. Sci. 258~(2) (2011) 743--747.

\bibitem{13spie88220S}
K.~Juodkazis, J.~Juodkazyt{\.e}, B.~{\v{S}}ebeka, S.~Juodkazis,
Reversible
  hydrogen evolution and oxidation mediated by molecular ion, in: SPIE Solar
  Energy+ Technology, International Society for Optics and Photonics, Proc.
  SPIE 8822, 2013, p. 88220S.

\bibitem{MesRu}
K.~Juodkazis, J.~Juodkazyt\.{e}, R.~Vilkauskait\.{e},
B.~\v{S}ebeka,
  V.~Jasulaitien\.{e}, Oxygen evolution on mixed ruthenium and nickel oxide
  electrode, Chemija 19 (2008) 1--6.

\bibitem{MesNi}
K.~Juodkazis, J.~Juodkazyt\.{e}, R.~Vilkauskait\.{e},
V.~Jasulaitien\.{e},
  Nickel surface anodic oxidation and electrocatalysis of oxygen evolution, J.
  Sol. State Electrochem. 12 (2014) 1469--1479.

\bibitem{YablonovitchSPIE}
O.~D. Miller, E.~Yablonovitch, Photon extraction: the key physics
for
  approaching solar cell efficiency limits, in: G.~S. Subramania,
  S.~Foteinopoulou (Eds.), Active Photonic Materials V, Proc. of SPIE vol.
  8808, 2013, p. 880807.

\bibitem{Bard}
A.~J. Bard, M.~A. Fox, Artificial photosynthesis: Solar splitting
of water to
  hydrogen and oxygen, Acc. Chem. Res. 28 (1995) 141--145.

\bibitem{Izumiya}
K.~Izumiya, E.~Akiyama, H.~Habazaki, N.~Kumagai, A.~Kawashima,
K.~Hashimoto,
  Anodically deposited manganese oxide and manganese - tungsten oxide
  electrodes for oxygen evolution from seawater, Electroch. Acta 43 (1998) 3303
  -- 3312.

\end{thebibliography}

\end{document}